\documentclass[aps,draft]{revtex4}
\usepackage[dvips]{graphicx}
\begin{document}

\title{Quantization of Longitudinal Electric Waves in Plasmas}
\author{ Levan N.Tsintsadze }
\thanks{Also at Department of Plasma Physics, E.Andronikashvii Institute of Physics, Tbilisi,
Georgia}
\affiliation{Graduate School of Science, Hiroshima University, Higashi-Hiroshima, Japan}

\date{\today}

\begin{abstract}
Effects of the Landau diamagnetism and the Pauli paramagnetism on the longitudinal electric wave characteristics in a quantum plasma are studied. It is shown that a dispersion relation of the longitudinal wave propagating along a magnetic field strongly depends on the magnetic field, in radical contrast to the classical case. New modes of quantum plasma waves due to the magnetic field are found.
\end{abstract}


\maketitle

\section{Introduction}

The influence of strong or superstrong magnetic field on the thermodynamic properties of medium and the propagation of proper waves is an important issue in supernovae and neutron stars, the convective zone of the sun, the early prestellar period of the evolution of the universe, as well as in the laboratory plasmas (the contemporary problems of laser-matter interaction).
Based on the astrophysical data, the surface magnetic field of a neutron star
is $H\sim 10^{11}-10^{13}G$, and the internal field can reach $H\sim 10^{15}G$ or
even higher \cite{lan}-\cite{lip}. As was shown by Bisnovati-Kogan \cite{bis}, the presence of rotation of stars may increase the magnetic field by an additional factor of $10^3-10^4$. In such strong magnetic fields, it is expected that the thermodynamic properties and wave dynamics in degenerate plasmas would be quite different governed by the quantum effects. This is true when the characteristic energy of electron on a Landau level reaches the nonrelativistic limit of the electron
chemical potential $\mu=\varepsilon_F=\frac{\hbar\mid e\mid H}{2m_ec},$ i.e.
$H=H_S\frac{v_F^2}{c^2}$, where $H_S=\frac{m_e^2c^3}{\mid e\mid \hbar}=
4.4\cdot 10^{13}G$ is the Schwinger's magnetic field, $v_F=p_F/m_e=(3\pi^2)^{1/3}\hbar n_e^{1/3}/m_e$ is the speed of electrons at the Fermi surface, $\hbar$ is the Planck constant divided by $2\pi$, $m_e$ is the electron rest mass, c is the speed of light in vacuum and $\mid e\mid$ is the magnitude of the electron charge.

As is well known, the strong magnetic field in the Fermion gas leads to two
magnetic effects \cite{land}. Namely, they are the Pauli paramagnetism due to the spin of electrons and
the Landau diamagnetism due to the quantization of the orbital motion of the electrons. We should note here that Klimontovich and
Silin \cite{kli} have derived the Wigner's type of non-relativistic quantum kinetic equation for a plasma, and in the linear approximation obtained dispersion relations for longitudinal plasma waves and transverse electromagnetic waves, but they did not take into account the quantization of the orbital motion of electrons and the spin of electrons.
It is also well known that in the classical limit the propagation of longitudinal wave along a magnetic field is unaffected by the magnetic field.

In this Letter, we show that the situation radically changes in a quantum plasma when the Landau diamagnetism and the Pauli paramagnetism are incorporated, by considering a special case of propagation of linear longitudinal waves along the homogeneous and time independent magnetic field.

\section{Some thermodynamical quantities}

We first calculate the thermodynamical quantities in the presence of a magnetic field. In a strong magnetic field H, as is known the motion in a plane perpendicular to the magnetic field is quantized \cite{land}, and the electron energy levels $\varepsilon_e^{\ell,\sigma}$ are determined in the  non-relativistic limit by the expression
\begin{eqnarray}
\label{eel}
\varepsilon_e^{\ell,\sigma}=\frac{p_z^2}{2m_e}+(2\ell+1+\sigma)\beta H\ ,
\end{eqnarray}
where $p_z$ is the electron momentum in the z-direction, $\beta=\frac{\mid e\mid\hbar}{2m_ec}$ is the Bohr magneton, $\sigma$ is the operator the z component of which describes the spin orientation $\vec{s}=\frac{1}{2}\vec{\sigma }$ ($\sigma=\pm 1$), and $\ell$ is the orbital quantum number ($\ell=0,1,2,...$).

From the expression (\ref{eel}) one sees that the energy spectrum of electrons consist of the lowest Landau level, $\ell=0$, $\sigma=-1$, and pairs of degenerate levels with opposite polarization, $\sigma=1$. Thus each value with $\ell\neq 0$ occurs twice, and that with $\ell=0$ once. Therefore, $\varepsilon_e^{\ell,\sigma}$can be rewritten as
\begin{eqnarray}
\label{reel}
\varepsilon_e^{\ell,\sigma}=\varepsilon_e^{\ell}=\frac{p_z^2}{2m_e}+2\ell\beta H=\frac{p_z^2}{2m_e}+\ell\hbar\omega_c \ ,
\end{eqnarray}
where $\omega_c=\frac{\mid e\mid H}{m_ec}$ is the cyclotron frequency of the electron.

The number of quantum states \cite{land} of a particle moving in a volume V and the interval $dp_z$ for any value of $\ell$ is
\begin{eqnarray}
\label{nqs}
\frac{2V\mid e\mid Hdp_z}{(2\pi\hbar)^2c}=\frac{V\varepsilon_{F\alpha}\eta_\alpha m_\alpha dp_z}{2\pi^2\hbar^3}\ ,
\end{eqnarray}
where suffix $\alpha$ stands for the particle
species, $\eta_\alpha=\frac{\hbar\omega_{c\alpha}}{\varepsilon_{F\alpha}}$ and $\varepsilon_{F\alpha}=\frac{p_{F\alpha}^2}{2m_\alpha}$. Note that for the degenerate particles $\eta_\alpha=\eta$ does not depend on the mass of particles.

The equilibrium total density of particles is defined as
\begin{eqnarray}
\label{tnd}
n_\alpha=\frac{m_\alpha\varepsilon_{F\alpha}\eta}{2\pi^2\hbar^3}\sum_{\ell=0}^\infty\int_{-\infty}^\infty dp_z \ f_\alpha(p_z,\ell) \ ,
\end{eqnarray}
where $f_\alpha(p_z,\ell)$ is the Fermi distribution function
\begin{eqnarray}
\label{fdf}
f_\alpha(p_z,\ell)=\frac{1}{\exp \left\{\frac{\frac{p_z^2}{2m_\alpha}+\ell\hbar\omega_{c\alpha}-\mu_\alpha}{T_\alpha}\right\}+1}\ ,
\end{eqnarray}
$\mu_\alpha$ being the chemical potential.

To evaluate the density $n_\alpha$ from the expression (\ref{tnd}), we shall consider a plasma at the temperature limit
$\mid \ell\hbar\omega_{c\alpha}-\mu_\alpha\mid\gg T_\alpha$. In this case the Fermi distribution function is in a good approximation described by the Heaviside step function $H(\mu_\alpha-\varepsilon_\alpha^\ell)$, which equals 1 for $\mu_\alpha=\varepsilon_{F\alpha}\geq\varepsilon_\alpha^\ell$ and zero for $\varepsilon_{F\alpha}<\varepsilon_\alpha^\ell$. We also note here that the maximum of $\ell$ should be $\ell_{max}=1/\eta$, since $p_z=p_{F\alpha}(1-\ell\eta)^{1/2}$ is real.

Let us recall some purely quantum mechanical features of a macroscopic system. It is well known that there is an extremely
high density of levels in the energy eigenvalue spectrum of a macroscopic system. We know also that the number of levels in a
given finite range of the energy spectrum of a macroscopic system increases exponentially with the number of particles N in the
system, and separations between levels are given by numbers of the $10^{-N}$. Therefore we can conclude that in such case the
spectrum is almost continuous, and a quasi-classical approximation is applicable. Thus, we can replace the summation in Eq.(\ref{tnd}) by an integration $(\sum_1^{\ell_{max}}\rightarrow\int_1^{\ell_{max}}d\ell)$ to obtain after a simple integration an expression of the density $n_\alpha$. The result is
\begin{eqnarray}
\label{fden}
n_\alpha=\frac{p_{F\alpha}^3}{2\pi^2\hbar^3}\left\{\eta+\frac{2}{3}(1-\eta)^{3/2}\right\}\ .
\end{eqnarray}
In equation (\ref{fden}) the first term is the contribution from the lowest Landau level ($\ell=0$), i.e. this term is associated with the Pauli paramagnetism and self-energy of particles. The second one results from the summation over all higher Landau levels.

If the magnetic field is absent $(\eta=0$), then for the density we get the well known expression
\begin{eqnarray}
\label{wde}
n_{\alpha_{\mid\eta=0}}=\frac{p_{F\alpha}^3}{3\pi^2\hbar^3}\ .
\end{eqnarray}
Whereas, if the magnetic field is very strong $\eta>1$, the sum in Eq.(\ref{tnd}) vanishes and all electrons are at the main level, which means that the gas is fully polarized and spins of all particles are aligned opposite to the magnetic field. Hence, in this case for the density we have
\begin{eqnarray}
\label{smd}
n_\alpha=\frac{p_{F\alpha}^3}{2\pi^2\hbar^3}\ \eta\ .
\end{eqnarray}
We now use Eq.(\ref{fden}) to write the limiting Fermi energy and the degenerate temperature as
\begin{eqnarray}
\label{lfe}
\varepsilon_{F\alpha}=T_{F\alpha}=\frac{p_{F\alpha}^2}{2m_\alpha}=\frac{(2\pi^2)^{2/3}\hbar^2n_\alpha^{2/3}}{2m_\alpha
\left\{\eta+\frac{2}{3}(1-\eta)^{3/2}\right\}^{2/3}}\ .
\end{eqnarray}

In the case when $\eta>1$, Eq.(\ref{lfe}) reduces to
\begin{eqnarray}
\label{rt}
T_{F\alpha}=\frac{(2\pi^2)^{2/3}\hbar^2}{2m_\alpha}\Bigl(\frac{n_\alpha}{\eta}\Bigr)^{2/3}\ ,
\end{eqnarray}
which reads that the degenerate temperature for the given $n_\alpha$ is reverse dependent on the magnetic field.

Noting the relation $\eta=\hbar\omega_c/\varepsilon_F$ equation (\ref{rt}) can be rewritten in the form
\begin{eqnarray}
\label{nrt}
T_{Fe}=\gamma \Bigl(\frac{n}{H}\Bigr)^2  \ ,
\end{eqnarray}
where
\begin{eqnarray*}
\gamma=\frac{\pi^4\hbar^4}{2m_e}\frac{c^2}{e^2} \ .
\end{eqnarray*}

The condition for an ideal gas is that the energy of the Coulomb interaction between electrons and ions is of the order of $Ze^2/r_0$, where $Ze$ is the ion charge and $r_0\sim (Z/n)^{1/3}$ is the mean distance between the electrons and ions, and this energy  should be small in comparison with the mean kinetic energy of the electrons (\ref{lfe}) or (\ref{rt}). In the case $\eta<1$, this condition $\varepsilon_F\gg Ze^2/r_0$ yields
\begin{eqnarray}
\label{c1}
n\gg\Bigl(\frac{2m_ee^2}{(2\pi)^{2/3}\hbar^2}\Bigr)^3Z^2\left\{\eta+\frac{2}{3}(1-\eta)^{3/2}\right\}^2 \ .
\end{eqnarray}
In the opposite case $\eta>1$, using Eq.(\ref{nrt}) we obtain
\begin{eqnarray}
\label{c2}
n\gg\Bigl(\frac{e^2}{\gamma}\Bigr)^{3/5}Z^{2/5}H^{6/5} \ .
\end{eqnarray}
Finally, in the case $\eta=0$ (H=0) we have \cite{land}
\begin{eqnarray*}
\label{c3}
n\gg\Bigl(\frac{2m_ee^2}{(3\pi^2)^{2/3}\hbar^2}\Bigr)^3Z^2 \ .
\end{eqnarray*}

\section{Longitudinal plasma waves}

We now consider a spectra of longitudinal waves of a strongly magnetized spatially homogeneous collisionless plasma in thermodynamic equilibrium, supposing that electrons are  degenerate, but not ions. Namely, the equilibrium distribution function for the electrons is assumed to be the step function (\ref{fden}) (in the limit $T_e=0$ or $T_F\gg T_e$). Whereas the ions are unmagnetized and the equilibrium distribution function for them is Maxwellian. As in the previous section, in the distribution function of electrons we take into account the quantization of the orbital motion of the electrons in the expression of energy.

The magnetic field is assumed to be directed along the z-axis and the longitudinal wave propagates in the same direction. Moreover, the amplitude of the longitudinal waves $E(z,t)=-\partial\varphi/\partial z=\delta E$ and a perturbation of the equilibrium distribution function of the electrons $f_{0e}^\ell$ ( $f_e=\sum_\ell f_e^\ell (z,t,p_z,\ell)$ ) and the ions $f_{0i}$ are small, i.e., $\mid\delta f_e^\ell\mid =\mid f_e^\ell-f_{0e}^\ell\mid\ll f_{0e}^\ell$ and $\mid\delta f_i\mid =\mid f_i-f_{0i}\mid\ll f_{0i}$.

In our previous paper \cite{tsin}, we have derived a new type of quantum kinetic equations of the Fermi particles. For our purpose, we employ a novel equation with quantum Madelung term obtained in \cite{tsin} neglecting particles collisions, which reads
\begin{eqnarray}
\label{tob}
\frac{\partial f_\alpha }{\partial t}+\left( \vec{v}\cdot\nabla \right) f_\alpha
+e_\alpha \Bigl(\vec{E}+\frac{\vec{v} \times \vec{H}}{c}\Bigr)\frac{\partial f_\alpha }{
\partial \vec{p}}+\frac{\hbar^2}{2m_\alpha }\nabla \frac{1}{\sqrt{
n_\alpha }}\Delta \sqrt{n_\alpha }\ \frac{\partial f_\alpha }{\partial \vec{p}}+\beta\nabla (\vec{\sigma}\cdot\vec{H})
\frac{\partial f_\alpha }{\partial \vec{p}}=0 \ .
\end{eqnarray}
Hereafter, we study the propagation of longitudinal plasma waves, but not spin waves. To this end, we linearize Eq.(\ref{tob}) for electrons, the kinetic equation of ions and the Poisson equation with respect to perturbations to get
\begin{eqnarray}
\label{per1}
\frac{\partial\delta f_e^\ell}{\partial t}+v\frac{\partial\delta f_e^\ell}{\partial z}+\Bigl(e\frac{\partial\varphi}{\partial z}+
\frac{\hbar^2}{4m_e}\frac{\partial^3}{\partial z^3}\frac{\delta n_e}{n_{0e}}\Bigr)\frac{\partial f_{0e}^\ell}{\partial p_z}=0
\end{eqnarray}
\begin{eqnarray}
\label{per2}
\frac{\partial\delta f_i}{\partial t}+v\frac{\partial\delta f_i}{\partial z}-e\frac{\partial\varphi}{\partial z}
\frac{\partial f_{0i}}{\partial p_z}=0
\end{eqnarray}
\begin{eqnarray}
\label{poi}
\frac{\partial^2\varphi}{\partial z^2}=4\pi e\left\{\frac{m_e\eta\varepsilon_F}{2\pi^2\hbar^3}\sum_{\ell=0}^\infty\int dp_z\delta
f_e^\ell-\int dp_z\delta f_i\right\} \ ,
\end{eqnarray}
where $n_{0e}$ is the equilibrium total number density of electrons.

We now look for wave solutions in space and time for $\delta f_e^\ell$, $\delta f_i$ and $\varphi$, assuming that they are proportional to $\exp{i(kz-\omega t)}$.

The expression of the electron density perturbation now is
\begin{eqnarray*}
\delta n_e=\frac{m_e\eta\varepsilon_F}{2\pi^2\hbar^3}\sum_{\ell=0}^\infty\int dv\delta f_e^\ell(k,\omega, v_z,\ell) \ .
\end{eqnarray*}
That is
\begin{eqnarray}
\label{nper}
\delta n_e=\frac{m_e\eta\varepsilon_F}{2\pi^2\hbar^3\Gamma_\ell}\sum_{\ell=0}^\infty\int\frac{dv\partial f_{0e}^\ell/\partial v}{
\omega-k v} \ ,
\end{eqnarray}
where $f_{0e}^\ell=H(\varepsilon_F-\varepsilon_e^\ell)$ is the Heaviside step function,
\begin{eqnarray}
\label{Gamma}
\Gamma_\ell=1+\frac{m_e^2\eta\varepsilon_F}{2\pi^2\hbar^3k}\frac{\omega_q^2}{n_{0e}}\sum_{\ell=0}^\infty\int\frac{dv\partial  f_{0e}^\ell/\partial v}{\omega-k v} \ ,
\end{eqnarray}
$\omega_q=\hbar k^2/2m_e$ being the frequency of quantum oscillations of electron, and
\begin{eqnarray}
\label{part}
\frac{\partial f_{0e}^\ell}{\partial v}=-m_ev\delta(\varepsilon_e^\ell-\varepsilon_F) \ ,
\end{eqnarray}
$\varepsilon_e^\ell$ being the electron energy levels in a magnetic field, the expression of which is given by Eq.(\ref{reel}).

For the ion density perturbation from Eq.(\ref{per2}) follows
\begin{eqnarray}
\label{idp}
\delta n_i=-\frac{e\varphi k}{m_i}\int\frac{dv\partial  f_{0i}/\partial v}{\omega-k v} \ ,
\end{eqnarray}
where $f_{0i}=n_{0i}\Bigl(\frac{m}{2\pi T_i}\Bigr)^{1/2}\exp\Bigl(-\frac{mv^2}{2T_i}\Bigr)$ is the Maxwellian distribution function.

Substituting Eqs.(\ref{nper}) and (\ref{idp}) into the Poisson equation (\ref{poi}) and integrating it over v, we obtain the dispersion relation for longitudinal oscillations in a partially degenerate plasma
\begin{eqnarray}
\label{disp}
\varepsilon(k,\omega)=1-\frac{4\pi e^2m_e\eta\varepsilon_Fv_F}{\Gamma_\ell k\pi^2\hbar^3}\left\{\frac{1}{\omega^2-k^2v_F^2}+
\sum_{\ell=1}^{\ell_{max}}\frac{\sqrt{1-\eta\ell}}{\omega^2-k^2v_F^2(1-\eta\ell)}\right\}+\frac{\omega_{pi}^2}{k^2v_{tri}^2}
\left\{1-I_+\Bigl(\frac{\omega}{kv_{tri}}\Bigr)\right\}=0 \ ,
\end{eqnarray}
where $\omega_{pi}$ is the ion plasma frequency, the function $I_+(x)=xe^{-x^2/2}\int_{i\infty}^xd\tau e^{\tau^2/2}$ has been studied in details in Ref.\cite{ale} and has such asymptotes
\begin{eqnarray*}
I_+(x)=1+\frac{1}{x^2}+\frac{3}{x^4}+...-i\sqrt{\frac{\pi}{2}}xe^{-x^2/2} \hspace{.7cm} if \hspace{.7cm} x\gg 1 \\
I_+(x)\simeq -i\sqrt{\frac{\pi}{2}}x \hspace{1cm} if \hspace{1cm} x\ll 1 \ .
\end{eqnarray*}

Replacing the summation in (\ref{disp}) by an integral, after a simple integration we finally obtain the desired dispersion equation
\begin{eqnarray}
\label{fdisp}
1-\frac{4\pi e^2m_e\eta\varepsilon_Fv_F}{\Gamma_\ell\pi^2\hbar^3}\left\{\frac{1}{\omega^2-k^2v_F^2}-\frac{2\sqrt{1-\eta}}{
\eta k^2v_F^2}\Bigl(1-\frac{\omega}{2kv_F\sqrt{1-\eta}}\ln\frac{\omega+kv_F\sqrt{1-\eta}}{\omega-kv_F\sqrt{1-\eta}}\Bigr)\right\}+
\nonumber \\
\frac{\omega_{pi}^2}{k^2v_{tri}^2}
\left\{1-I_+\Bigl(\frac{\omega}{kv_{tri}}\Bigr)\right\}=0
\end{eqnarray}
and
\begin{eqnarray}
\label{fgamma}
\Gamma_\ell=1-\frac{p_F^3\omega_q^2\eta}{2\pi^2\hbar^3n_{0e}}\left\{\frac{1}{\omega^2-k^2v_F^2}-\frac{2\sqrt{1-\eta}}{
\eta k^2v_F^2}\Bigl(1-\frac{\omega}{2kv_F\sqrt{1-\eta}}\ln\frac{\omega+kv_F\sqrt{1-\eta}}{\omega-kv_F\sqrt{1-\eta}}\Bigr)\right\} \ .
\end{eqnarray}

In the absence of the external magnetic field, from Eq.(\ref{fdisp}) we can immediately recover the result (Eq.(14)) of Ref.\cite{tsin}. It should be emphasized that the dispersion equation (\ref{fdisp}) is general (in the one-dimensional case), which describes high-frequency oscillations of electrons, as well as branch of ion spectrum.

Let us now discuss some special cases. We first consider a high-frequency longitudinal waves neglecting the contribution of ions, and show that the magnetic field is responsible for new modes of quantum plasma waves.

In the case of strong magnetic field, i.e. $\eta >1$, the sum in Eq.(\ref{disp}) vanishes and all electrons are at the Landau ground level ($\ell=0$), which means that the electron gas is fully polarized. Using the expression of the electron density (\ref{smd}), from Eq.(\ref{disp}) we then get the dispersion equation
\begin{eqnarray}
\label{smdis}
\omega^2=\omega_p^2+\omega_q^2+k^2v_F^2  \ ,
\end{eqnarray}
where $\omega_p=\sqrt{4\pi e^2n_{0e}/m_e}$ and $n_{0e}$ is defined by Eq.(\ref{smd}).
We specifically note here that all terms in Eq.(\ref{smdis}) can be the same order, or even $kv_F>\omega_p$. It should be also emphasized that this spectra (\ref{smdis}) is a new branch of frequencies due to the quantization of the orbital momentum of electrons.

Next in the case when $\hbar\omega_{ce}<\varepsilon_F=p_F^2/2m_e$, i.e. $\eta<1$, noting that in the range of frequencies $\omega>kv_F\sqrt{1-\eta}$ the roots of Eq.(\ref{fdisp}) are real, we obtain the spectra of quantum Langmuir electron waves in the presence of magnetic field
\begin{eqnarray}
\label{lang}
\omega^2=\omega_p^2+\omega_q^2+k^2v_F^2\frac{1}{1+\frac{3}{4}\lambda_B^2\lambda_{TF}^2k^4+3k^2\lambda_{TF}}\left\{
3k^2\lambda_{TF}^2+\frac{\eta(1+\frac{3}{4}\lambda_B^2\lambda_{TF}^2k^4)}{\eta+\frac{2}{3}(1-\eta)^{3/2}}\right\} \ .
\end{eqnarray}
Here we have introduced the Thomas-Fermi screening length $\lambda_{TF}=v_F/\sqrt{3}\omega_p$ and de Broglie wavelength $\lambda_B=\hbar/p_F$.

As is well known in the classical plasma at the electron temperature larger than that of the ions a slowly damping ion waves and the Langmuir electron waves can simultaneously propagate. In the quantum plasma one should also expect the existence of the ion waves. In order to demonstrate this, we consider an intermediate wave range where the phase velocity satisfies the inequality
\begin{eqnarray}
\label{ine}
v_{tri}\ll\frac{\omega}{k}\ll v_F\sqrt{1-\eta} \ .
\end{eqnarray}
In this case one assumes that the characteristic dimension R of inhomogeneities in the plasma is larger than the electron Thomas-Fermi length $\lambda_{TF}$. So, for $\lambda_{TF}\ll R$ we further assume that the quasi neutrality
\begin{eqnarray}
\label{qneu}
n_e=n_i
\end{eqnarray}
is satisfied, and we treat ions as cold. With this assumption Eq.(\ref{fdisp}) admits the complex roots ($\omega=\omega^\prime+i\omega^{\prime\prime}$) such as
\begin{eqnarray}
\label{comp1}
\omega^{\prime 2}=\Bigl(\eta+\frac{2}{3}(1-\eta)^{3/2}\Bigr)\left\{\frac{m_e}{m_i}\frac{k^2v_F^2}{\eta+2\sqrt{1-\eta}}+\frac{3}{2}
\frac{\hbar^2k^4}{4m_im_e}\right\}
\end{eqnarray}
and the imaginary part
\begin{eqnarray}
\label{comp2}
\omega^{\prime\prime}=-\frac{\pi}{4}\frac{m_e}{m_i}kv_F\ \frac{\eta+\frac{2}{3}(1-\eta)^{3/2}}{\eta+2\sqrt{1-\eta}} \ .
\end{eqnarray}
We note here that in the absence of the magnetic field $\eta=0$ the expressions (\ref{comp1}) and (\ref{comp2}) converge to the equations (26) and (27) of Ref.\cite{tsin}.

We now study the case when the phase velocity of low frequency waves satisfies the following conditions
\begin{eqnarray}
\label{cond}
v_{tri}\ll\frac{\omega}{k}\ll v_F \hspace{.7cm} and \hspace{.7cm} \frac{\omega}{k}\gg v_F\sqrt{1-\eta} \ .
\end{eqnarray}
The second inequality of (\ref{cond}) indicates that the cyclotron energy $\hbar\omega_{ce}$ is rather close to the limiting Fermi energy $\varepsilon_F$, or $\eta\sim 1$. This branch of the oscillation spectrum has very weak damping when the ion temperature is small in comparison with the electron degeneracy temperature
\begin{eqnarray}
\label{temp}
T_i < T_F \ .
\end{eqnarray}
With the inequalities (\ref{cond}) at hand, from the dispersion relation (\ref{fdisp}) for the longest waves $k\lambda_{TF}\ll 1$ we obtain the real
\begin{eqnarray}
\label{real}
\omega^{\prime 2}=\frac{m_e}{m_i}\left\{\frac{\eta+\frac{2}{3}(1-\eta)^{3/2}}{\eta}\ k^2v_F^2+\omega_q^2\right\}
\end{eqnarray}
and the imaginary part of $\omega$, which defines the damping rate
\begin{eqnarray}
\label{imag}
\omega^{\prime\prime}=-\sqrt{\frac{\pi}{8}}\ \frac{\omega^{\prime 4}}{k^3v_{tri}^3}\exp\Bigl(-\frac{\omega^2}{2k^2v_{tri}^2}\Bigr)  \ .
\end{eqnarray}
Note that in this case  the contribution of electrons to $\omega^{\prime\prime}$ is zero. So that the damping is due to the ions alone.

It should be emphasized that the expressions (\ref{real}) and (\ref{imag}) are novel, and exist only if one takes into account the Landau quantization.

\section{Summary}

We have investigated the effects of the quantization of the orbital motion of electrons and the spin of electrons on the propagation of longitudinal waves in a quantum plasma. It is well known that in the classical approximation the propagation of longitudinal wave along a magnetic field is unaffected by the magnetic field. We found that the situation radically changes in a quantum plasma when the Landau diamagnetism and the Pauli paramagnetism are taken into account. Namely, we derived a novel dispersion equation of the longitudinal wave propagating along the homogeneous and time independent magnetic field, which contains all the information on the quantum effects and exhibits the strong dependence on the magnetic field. Studying this dispersion relation we have disclosed new modes of quantum plasma waves. These investigations may play an essential role for the description of complex phenomena that appear in dense astrophysical objects, as well as in the next generation intense laser-solid density plasma experiments.


\begin{references}

\bibitem{lan} J.Landstreet,  Phys. Rev. {\bf 153}, 1372 (1967).

\bibitem{sha} S.L.Shapiro and S.A.Teukolsky, {\sl Black Holes, White Dwarfs, and Neutron Stars} (John Wiley and Sons, New York, 1981).

\bibitem{lip} V.M.Lipunov, {\sl Neutron Star Astrophysics} (Nauka, Moscow, 1987).

\bibitem{bis} G.S.Bisnovati-Kogan, Astron. Zh. {\bf 47}, 813 (1970).

\bibitem{land} L.D.Landau and E.M.Lifshitz, {\sl Statistical Physics}, Part 1 (Butterworth-Heinemann, Oxford, 1998).

\bibitem{wig} E.P.Wigner, Phys. Rev. {\bf 40}, 749 (1932).

\bibitem{kli} Yu.L.Klimontovich and V.P.Silin, Zh. Eksp. Teor. Fiz. {\bf 23}, 151 (1952).

\bibitem{tsin} N.L.Tsintsadze and L.N.Tsintsadze, in {\sl From Leonardo to ITER: Nonlinear and Coherence Aspects},
edited by Jan Weiland, AIP Proc. No. CP1177 (AIP, New York, 2009), 18; e-print arXiv: physics/0903.5368v1.

\bibitem{ale} A.F.Alexandrov, L.S.Bogdankevich and A.A.Rukhadze, {\sl Principals of Plasma Electrodynamics} (Springer, Heidelberg, 1984).

\end{references}
\end{document}